\documentclass{optica-article}

\journal{opticajournal} 

\articletype{Research Article}

\usepackage{lineno}

\begin{document}

\title{A New Perspective on Matrix Representation of Paraxial Geometric Optics using Two Kinds of Three-Matrix Decompositions of the $2\times 2$ Special-Linear-Group Matrices}

\author{Satoshi Itoh \authormark{1}}

\address{\authormark{1} Osaka university, Faculty of earth and space science, 1-1, Machikaneyama-cho, Toyonaka, Osaka, Japan, 560-0043\\}
\email{\authormark{ito\_s@ess.sci.osaka-u.ac.jp}  \authormark{itohsatoshi@aol.com}} 


\begin{abstract*} 
We require decomposition methods for the ABCD-matrix formulation in rotationally symmetric paraxial geometric optics when designing a multi-component optical system from a given single paraxial specification (represented by an ABCD matrix) to optimize non-paraxial specifications (e.g., optical aberrations).
In this study, we focus on two kinds of three-matrix decomposition of ABCD matrices to formulate a transformation between the two kinds of decomposition, which can increase or decrease the number of refractive surfaces in the optical configuration while keeping the paraxial specifications fixed. This transformation is not a parameterization of ABCD matrices nor a mapping between existing parameterizations, but a canonical structural relation derived from the canonical three-matrix decompositions. This approach is useful for the optical design of multi-component systems with optimized non-paraxial characteristics.
\end{abstract*}

\section{Introduction}
In the field of geometric optics, the first-order approximation (paraxial approximation or Gaussian optics) provides us with a global picture of the optical configuration.  
Thanks to the first-order (linear) approximation, we can use linear algebra to express optical propagation\cite{hecht2017optics}; in other words, a matrix can represent ray response in the paraxial regime.
When the optical configuration has rotational symmetry, the ray vector needs only two components: ray height and direction. Thus, the propagation matrices are $2\times 2$ real matrices.
From the law of conservation of radiance\cite{nicodemus1963radiance}, the $2\times 2$ propagation matrices must satisfy the condition that the determinant of the matrix is unity (in a certain convention), which means the matrices express the members of the special linear group $\mathrm{SL}(2,\mathbb{R})$\cite{hall2003lie}.

In a purely mathematical perspective, we can express the $\mathrm{SL}(2,\mathbb{R})$ matrices using the following parametrization (often referred to as ABCD matrices):
\begin{equation}\mathrm{SL}(2,\mathbb{R}) =\left\lbrace
    \begin{pmatrix}
   A & B \\
   C & D
\end{pmatrix}\
\middle|\  A,B,C,D \in \mathbb{R}, \mathrm{and}\  AD-BC=1 
\right\rbrace.
\end{equation}
We can express every rotationally symmetric multi-component configuration (e.g., multi-lens optics) using a single ABCD matrix in the first-order approximation.
This formulation is convenient for observing the response of the given optical system as a single matrix.

On the other hand, the ABCD-matrix formulation itself is insufficient for us to design a multi-component configuration from a given single paraxial specification (represented by a given ABCD matrix) to optimize some non-paraxial specifications.
In this case, we need a formula to decompose the ABCD matrices into various patterns of products of some elementary matrices.
The way for the decomposition of the ABCD matrix exists infinitely and several studies on this topic exist\cite{2008ApOpt..47E..88L,2014JMOp...61..161B,Barion:24}.
In this study, we focus on only two fundamental kinds of three-matrix decomposition of ABCD matrices to formulate a transform between the two kinds of decomposition.
This transform provides us with a new perspective to design a multi-component optical configuration from a given single paraxial specification.
Although matrix decompositions of ABCD matrices have been studied in both the optics and mathematics communities, the potential connections between paraxial geometric optics and algebraic group theory—specifically Chevalley group theory—have not been widely explored.
This work aims to bridge this gap by introducing two canonical three-matrix decompositions and clarifying the transformation structure between them.
In Section \ref{secP}, we defined the symbols for the present study.
Section \ref{secT} formulates the two kinds of matrix decompositions of the $\mathrm{SL}(2,\mathbb{R})$ matrices. 
Section \ref{secT2} investigates the transformation between the two kinds of decompositions for a single ABCD matrix and discusses its application.
Section \ref{UC} exhibits a use case of the proposed concept.
In Section \ref{secC}, we compile the findings in this study.

\section{Preparation\label{secP}}
Here, we define symbol notation in the matrix representation of the rotationally symmetric paraxial optics, such that we can easily observe the mathematical structure in the theory. 
Almost all the conventions defined in Section \ref{secP} follow the notation in \cite{hecht2017optics}.
In this study, we define ray vectors $\mathbf{v}$ as follows:
\begin{equation}
    \mathbf{v}=    \begin{pmatrix}
   n \theta \\
   y
\end{pmatrix}, 
\end{equation}
where the symbols $n$, $\theta$, and $y$ denote the refractive index at the ray's location, the angle between the ray direction and the optical axis, and the ray height from the optical axis, respectively.

The transition matrix $T(p)$ is an operator that acts on the ray vectors, defined as follows:
\begin{equation}
    T(p)=    \begin{pmatrix}
   1&0 \\
   p&1
\end{pmatrix},\label{eq3}
\end{equation}
where $p=d/n_{d}$; the symbols $d$ and $n_{d}$ mean the distance of the transition and the refractive index within the transition interval, respectively. 
The following refraction matrix $R(q)$ is also an operator for the ray vectors:
\begin{equation}
    R(q)=    \begin{pmatrix}
   1&q \\
   0&1
\end{pmatrix},\label{eq4}
\end{equation}
where $q=-\mathcal{D}$; the symbol $\mathcal{D}$ indicates the focusing power of the refractive plane. 
The focusing power of the refractive plane $\mathcal{D}$ has the following definition:
\begin{equation}
    \mathcal{D}=\frac{n_{\mathrm{a}}-n_{\mathrm{b}}}{r},
\end{equation}
where the symbols ($n_{\mathrm{a}},n_{\mathrm{b}}$) mean the refractive indices (after, before) the refractive surface; the symbol $r$ expresses the radius of curvature of the refractive surface. 
When we utilize the concept ot thin lens for this refractive surface, we can define the focal length $f$ of this surface as follows:
\begin{equation}
    f=\frac{1}{D}.
\end{equation}

\section{Two Kinds of Three-Matrix Decompositions of the 2$\times$2 Special-Linear-Group Matrices\label{secT}}
Now, we formulate two types of three-matrix decompositions of $\mathrm{SL}(2,\mathbb{R})$ matrices: + (cross) decomposition (Section \ref{p}) and H composition (Section \ref{h}). 
We can interpret these decompositions in terms of the Chevalley group\cite{BB22782442} structure of $SL(2)$.
From the viewpoint of algebraic group theory, $SL(2,\mathrm{R})$ is a rank‑one Chevalley group\cite{BB22782442} associated with the root system $A_{1}$, where the symbol $\mathrm{R}$ means any commutative ring that includes the real field $\mathbb{R}$ as an example.
The optical case corresponds to the specialization $\mathrm{R}=\mathbb{R}$.
A fundamental property of this structure is that the group is generated (decomposed) entirely by its two root subgroups
\begin{equation}
X_{+}(t)=
\left\lbrace\left.\begin{pmatrix}
1 & t \\
0 & 1
\end{pmatrix}\right|t \in \mathrm{R}
\right\rbrace
\end{equation}
\begin{equation}
X_{-}(t)=
\left\lbrace\left.\begin{pmatrix}
1 & 0 \\
t & 1
\end{pmatrix}\right|t \in \mathrm{R}
\right\rbrace    
\end{equation}

corresponding to the positive and negative roots. These subgroups provide a canonical and non‑arbitrary generating set for $SL(2,\mathrm{R})$.
As stated in Lemma 26 in \cite{BB22782442}, the Chevalley group $G$ is generated by the root subgroups $F_{\alpha}$ and the Weyl elements $w_{\alpha}$ associated with simple roots $\alpha$. 
Since the Weyl group generated by these $w_{\alpha}$ acts transitively on the set of all roots, this is equivalent to saying that $G$ is generated by the root subgroups corresponding to \emph{all} roots.

In the context of paraxial optics, these two generators have direct physical interpretations: $X_{+}(t)$ corresponds to refraction at an interface, while $X_{-}(t)$ corresponds to free‑space propagation. Thus, the two elementary transformations of first‑order optics arise naturally from the Chevalley group structure. This perspective might clarify why the factorizations focused on in the present study simplify the paraxial geometrical optics.

Some of the $\mathrm{SL}(2,\mathbb{R})$ matrices, which express configurations important in the optical design, are decomposable with no + or H decomposition, but these matrices can be expressed as a product of two matrices decomposable with + and H decomposition (Section \ref{i}).

\subsection{+ (Cross) Decomposition\label{p}}
We assume an $\mathrm{SL}(2,\mathbb{R})$ matrix $M$:
\begin{equation}
    M=\begin{pmatrix}
   A & B \\
   C & D
\end{pmatrix},\label{eq6}
\end{equation}
where $AD-BC=1$.
When $B\neq 0$ or when the ABCD matrix $M$ is a transition matrix, we can decompose an $\mathrm{SL}(2,\mathbb{R})$ matrix $M$ as follows:
\begin{eqnarray}
    M&=& T(c)R(\beta)T(a).
\end{eqnarray}
where
\begin{equation}
    \begin{pmatrix}
   A & B \\
   C & D
\end{pmatrix}=\begin{pmatrix}
   1+a\beta & \beta \\
   a+c+ac\beta & 1+c\beta
\end{pmatrix}.\label{qwe}
\end{equation}
Note that we chose the Greek letter $\beta$ for the parameter of the refraction matrix, while using the Latin letters $a$ and $c$ for the transition matrix.
When $B\neq 0$, we can rewite Eq. (\ref{qwe}) as follows:
\begin{equation}
    a = \frac{A-1}{B},
\end{equation}
\begin{equation}
    \beta = B,
\end{equation}
and
\begin{equation}
    c = \frac{D-1}{B}.
\end{equation}
When the ABCD matrix $M$ is a transition matrix, we can rewite Eq. (\ref{qwe}) as follows:
\begin{equation}
    a+c=C,
\end{equation}
\begin{equation}
    \beta=0.
\end{equation}

\begin{figure}[htb]
    \centering
    \includegraphics[width=0.5\linewidth]{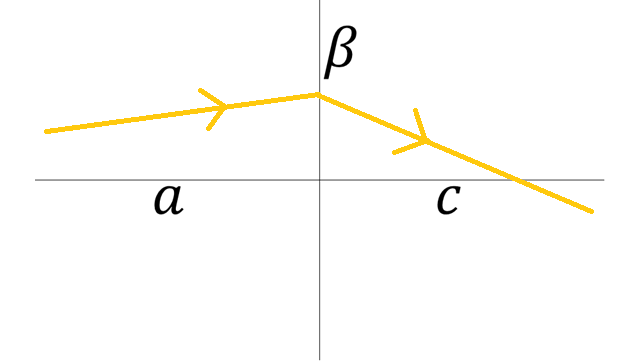}
    \caption{Schematic of the + decomposition $M=T(c)R(\beta)T(a)$. This diagram conceptually shows an optical configuration. In this diagram, we show an example ray that goes from left to right as the yellow line with arrows. The horizontal gray lines labeled by $a$ and $c$ express the light transition. The vertical gray line indicates refraction. The symbols $\beta$, $a$, and $c$ denote the parameters of the operator matrices.}
    \label{fig:placeholder1}
\end{figure}

This decomposition consists only of transition and refraction matrices.
Thus, we can describe the corresponding optical configuration as shown in Fig. \ref{fig:placeholder1}.   
Because the diagram shape of Fig. \ref{fig:placeholder1} looks like the + (cross) character, we refer to this decomposition as + (cross) decomposition in this study. 

\subsection{H Decomposition\label{h}}
\begin{figure}[htb]
    \centering
    \includegraphics[width=0.5\linewidth]{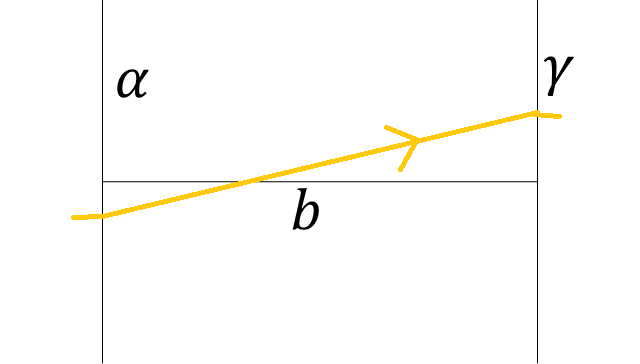}
    \caption{Schematic of the H decomposition $M=R(\gamma)T(b)R(\alpha)$. This diagram conceptually shows an optical configuration. In this diagram, we show an example ray that goes from left to right as the yellow line with arrows. The horizontal line labeled by $b$ expresses the light transition. The vertical lines labeled by $\alpha$ and $\gamma$ indicate refraction. The symbols $b$, $\alpha$, and $\gamma$ denote the parameters of the operator matrices.}
    \label{fig:placeholder2}
\end{figure}

When $C\neq 0$ or when the ABCD matrix $M$ is a refraction matrix, we can decompose the matrix $M$ as follows:
\begin{eqnarray}
    M&=& R(\gamma)T(b)R(\alpha).
\end{eqnarray}
where
\begin{equation}
    \begin{pmatrix}
   A & B \\
   C & D
\end{pmatrix}=\begin{pmatrix}
   1+\gamma b & \alpha +\gamma+\alpha \gamma b \\
   b & 1+\alpha b
\end{pmatrix}.\label{poi}
\end{equation}
Note that, in the same manner as Eq. (\ref{qwe}), we choice the Greek letters $\alpha$ and $\gamma$ for the parameter of the refraction matrices, while using the Latin letter $b$ for the transition matrix.
When $C\neq 0$, we can rewite Eq. (\ref{poi}) as follows:
\begin{equation}
    \alpha = \frac{D-1}{C},
\end{equation}
\begin{equation}
    b = C,
\end{equation}
and
\begin{equation}
    \gamma = \frac{A-1}{C}.
\end{equation}
When the ABCD matrix $M$ is a reflection matrix, we can rewite Eq. (\ref{qwe}) as follows:
\begin{equation}
    \alpha+\gamma=B,
\end{equation}
\begin{equation}
    b=0.
\end{equation}
Fig. \ref{fig:placeholder2} indicates the optical configuration corresponding to this decomposition.  
The diagram in Fig. \ref{fig:placeholder2}  has the shape of the character of H; we refer to this decomposition as the H decomposition in this study.

\subsection{$\mathrm{SL}(2,\mathbb{R})$ Matrices Decomposable with no + or H Decomposition\label{i}}
When $B=C=0$ in the ABCD matrix $M$, $M$ is decomposable with no + or H decomposition.
Since $AD-BC=1$, we can write the matrix $M$ in this case as follows:
\begin{equation}
    M=\begin{pmatrix}
   t^{-1} & 0 \\
   0 & t
\end{pmatrix},\label{eq11}
\end{equation}
where $t \in \mathbb{R}$ $(t\neq 0)$.
This matrix propagates all rays with zero height ($y=0$) to rays with zero height. 
In addition, this matrix brings all the rays with zero angles ($n\theta=0$) to rays with zero angles.
Thus, the matrix of Eq. (\ref{eq11}) expresses a special type of optical conjugate (imaging with a magnification ratio of $t$), where initial-plane rays parallel to the optical axes remain parallel to the optical axes in the final plane.

This type of optical configuration is important for building a telecentric imaging system.
Hence, decomposing these types of matrices is critical for a certain optical design. 
We can decompose this type of matrix into a product of two matrices decomposable with + or H decomposition.
To observe this fact, we consider matrices $N(f)$ decomposable with the + decomposition and H decomposition as follows:
\begin{eqnarray}
N(f)&=&T(f)R(-1/f)T(f) \nonumber\\
&=&R(-1/f)T(f)R(-1/f) \nonumber\\
&=& \begin{pmatrix}
   0 & -\frac{1}{f} \\
   f & 0
\end{pmatrix},\label{dhi}
\end{eqnarray}
where $f\in\mathbb{R}$ $(f\neq 0)$.
When we consider the refraction $R(-1/f)$ as a thin lens, the parameter $f$ means the focal length of the thin lens.
The matrices $N(f)$ represent an optical transformation such that the two components (position and direction) of a ray vector interchange and scale with factors determined by the focal length $f$.
In wave optics, Fourier optics works for this type of optical configuration.
In the mathematics of the Chevalley Group, $N(1)$ and $N(-1)$ correspond to the Weyl elements.
Using Eq. (\ref{dhi}), we can compose matrices $M$ in the form of Eq. (\ref{eq11}) as a tandem configuration of two thin lenses with focal lengths $f$ and $f'$ (telecentric imaging optics with a magnification factor $f'/f$):

\begin{eqnarray}
M=N(f)N(f')
&=& \begin{pmatrix}
   -\frac{f}{f'} & 0 \\
   0 & -\frac{f'}{f}
\end{pmatrix}.\label{uio}
\end{eqnarray}
Hence, ABCD matrices decomposable with no + or H decomposition are decomposable into two matrices decomposable with + and H decomposition.

\section{Transformation between + and H Decompostion\label{secT2}}
\begin{figure}[htb]
    \centering
    \includegraphics[width=0.8\linewidth]{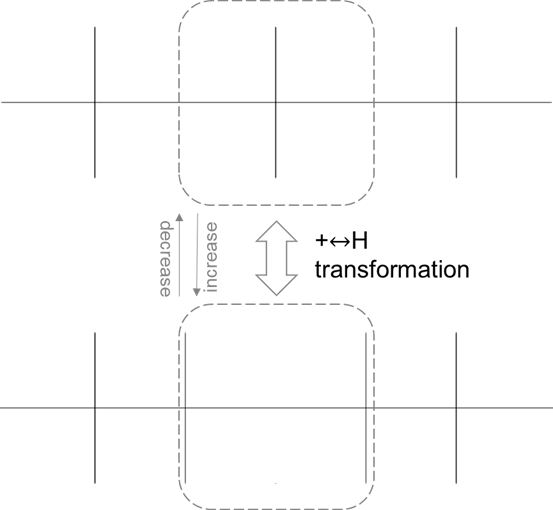}
    \caption{Schematic for how the + $\leftrightarrow$ H (cross $\leftrightarrow$ H) transformation increases or decreases the number of refractive surfaces in an optical configuration. These diagrams conceptually show optical configurations. In these diagrams, the ray goes from right to left. The horizontal lines express the light transition. The vertical line indicates refraction. The + $\leftrightarrow$ H transformation interchanges the + part within the dashed closed curve in the top panel and the H part within the dashed closed curve in the bottom panel.  }
    \label{fig:placeholder3}
\end{figure}

Here, we consider the transformation between the + and H decompositions of a single ABCD matrix. 
The transformation introduced here is not a parameterization of ABCD matrices, nor is it a known mapping between existing parameterizations.
It is defined as a structural transformation between two canonical three-matrix decompositions derived from Chevalley group theory.
This transformation is useful to increase or decrease the number of refractive surfaces in the optical configuration while keeping paraxial performance (Fig. \ref{fig:placeholder3}).
We can define this transformation as the following mapping $g$: $(a,\beta,c)\to(\alpha,b,\gamma)$ and its inverse mapping $g^{-1}$: $ (\alpha,b,\gamma)\to (a,\beta,c)$ that satisfy the following equation:
\begin{equation}
    T(c)R(\beta)T(a)= R(\gamma)T(b)R(\alpha).\label{ph}
\end{equation}

This transformation has two fundamental invariants $I$ and $J$ as follows; here, we define the term “invariant” to refer to a function of a given variable whose functional form (not only the value) remains unchanged even when subjected to a specified variable transformation:
\begin{equation}
    I=a \beta^2 c = \alpha b^2 \gamma
\end{equation}
\begin{equation}
    J=(a+c)\beta =(\alpha+\gamma)b.
\end{equation}
The conserved polynomials $I$ and $J$ are symmetric polynomials of the parameters appearing in the canonical three-matrix decompositions, and their invariance is defined with respect to the +/H transformation.
Therefore, they represent structural invariants of the decomposition itself rather than functions of the matrix elements $A$ and $D$.

When we compare the diagonal components of Eq. (\ref{qwe}) and(\ref{poi}), we observe that, to satisfy Eq. (\ref{ph}), we at least require the followings:
\begin{equation}
    a\beta = \gamma b.
\end{equation}
and
\begin{equation}
    c\beta = \alpha b
\end{equation}
Defining $\xi=a\beta = \gamma b$ and $\eta=c\beta = \alpha b$ and making elementary symmetric polynomials of the dimensionless quantities $\xi$ and $\eta$ leads to the expression of the above invariants $I$ and $J$:
\begin{equation}
    I=\xi\eta
\end{equation}
and
\begin{equation}
    J=\xi+\eta.
\end{equation}
Note that the value of $J+2$ is equal to the trace of the ABCD matrix; the value of $J$ corresponds to the trace of the difference between the ABCD matrix and the identity matrix.

Using these invariants $I$ and $J$, we can formulate the transformation as follows:
\begin{equation}
    \beta b =I+J\label{eq17}
\end{equation}
and
\begin{equation}
    a \alpha =c\gamma = \frac{I}{I+J}.\label{eq18}
\end{equation}
Hence, we can interpret 
the + $\leftrightarrow$ H transformation as swapping the transition and refraction matrices as follows:
\begin{eqnarray}
    T(a)&\leftrightarrow&R(\alpha)\nonumber\\
    R(\beta)&\leftrightarrow&T(b)\nonumber\\
    T(c)&\leftrightarrow&R(\gamma),
\end{eqnarray}
where the parameters $a,\beta,c,\alpha,b,$ and $\gamma$ satisfy the conditions of Eq. (\ref{eq17}) and Eq. (\ref{eq18}).

\section{An Use Case\label{UC}}
\begin{figure}[h]
    \centering
    \includegraphics[width=1.0\linewidth]{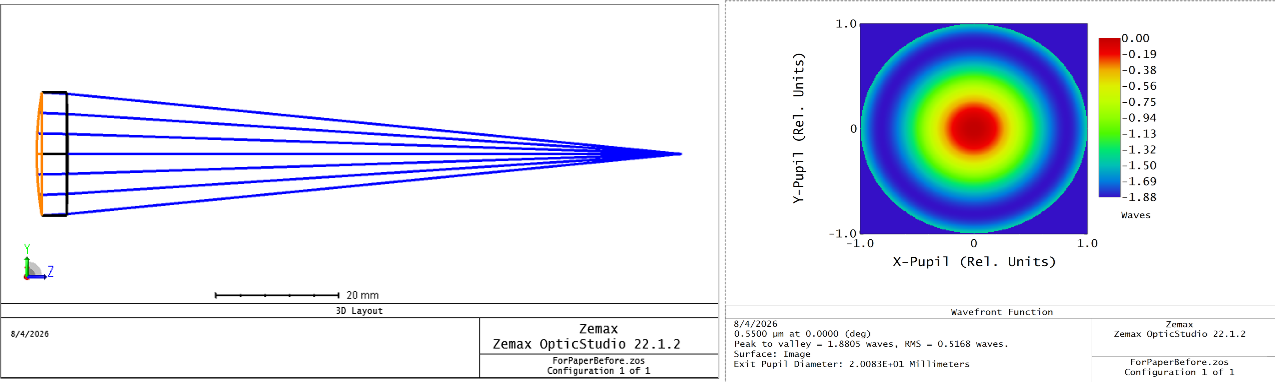}
    \caption{The initial configuration to be modified with the + $\leftrightarrow$ H transformation. The left panel shows the optical layout. The right panel shows the wavefront error of this configuration in the case of collimated light injection with the same direction as the optical axis. The assumed wavelength of light is $0.55\mathrm{\mu m}$.}
    \label{fig:e1}
\end{figure}

Here, we observe how to use the + $\leftrightarrow$ H transformation as a constraint in the lens-design optimization.
We modify an example lens configuration using the + $\leftrightarrow$ H transformation as a constraint in the optimization. 
We assume the initial lens design (Fig. \ref{fig:e1}, Table \ref{t1}, and Table \ref{tz1}) to be modified.
The initial lens design has a single lens with two refractive spherical surfaces (front and back surfaces of the lens). 

\begin{table}[h]
\caption{The surface parameters of the initial configuration to be modified with the + $\leftrightarrow$ H transformation (Fig. \ref{fig:e1}). The second column shows the physical thickness between the surface and the next surface. The third column indicates the refractive index of the medium filling the space between the surface and the next surface.}
  \label{tab:shape-functions}
  \centering
\begin{tabular}{cccc}
\hline
Surface \# &Curvature Radius&Thickness& Refractive Index \\
\hline
1&61.441 mm&5 mm&1.5185\\
2 &-413.373 mm&100 mm &1.000 \\
\hline
\end{tabular}
\label{t1}
\end{table}

\begin{table}[h]
\caption{The Zernike (standard) coefficients up to the 37th order for the initial configuration. The symbol $\rho$ indicates the pupil radial coordinate normalized by the physical radius of the pupil. The symbol $\lambda$ means the wavelength of light; $\lambda=0.55\mathrm{\mu m}$ here. We omitted the orders with exactly zero values of the coefficients.}
  \label{tab:shape-functions}
  \centering
\begin{tabular}{ccc}
\hline
Order&Base Function&Value of Coefficient\\
\hline
4&$\sqrt{3}(2\rho^2-1)$&-0.4108$\lambda$\\
11&$\sqrt{5}(6\rho^4-6\rho^2+1)$&0.3132$\lambda$\\
22&$\sqrt{7}(20\rho^6-30\rho^4+12\rho^2-1)$&0.0012$\lambda$\\
37&$\sqrt{9}(70\rho^8-140\rho^6+90\rho^4-20\rho^2+1)$&0.0000$\lambda$\\
\hline
\end{tabular}
\label{tz1}
\end{table}

\begin{figure}[h]
    \centering
    \includegraphics[width=0.6\linewidth]{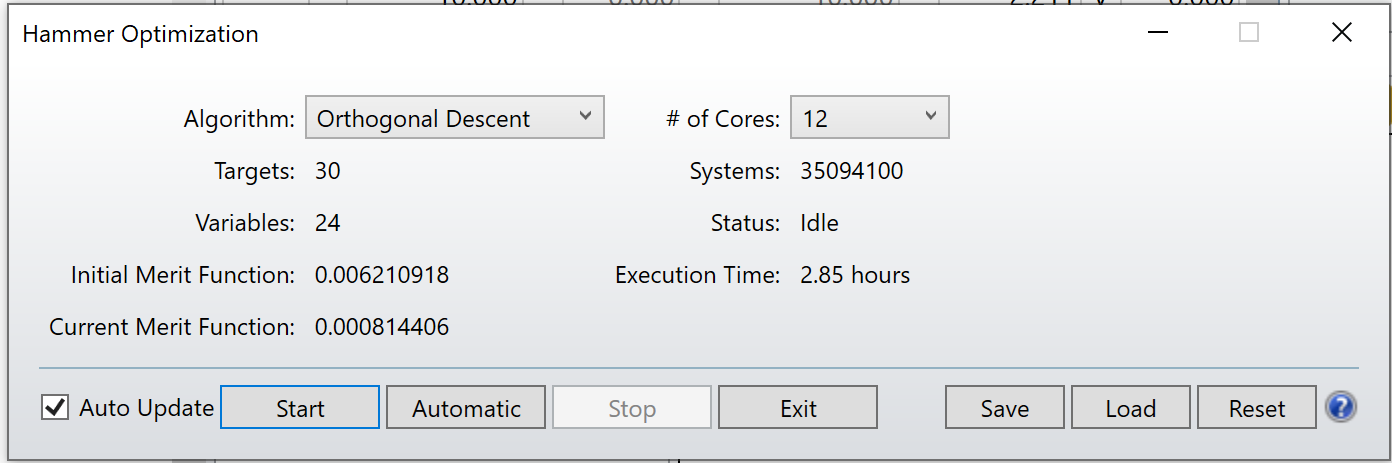}
    \caption{The screenshot of the optimization tool GUI after the optimization.}
    \label{fig:e5}

\end{figure}

We can compile the sequence of the design modification as follows.
\begin{figure}[h]
    \centering
\includegraphics[width=0.4\linewidth]{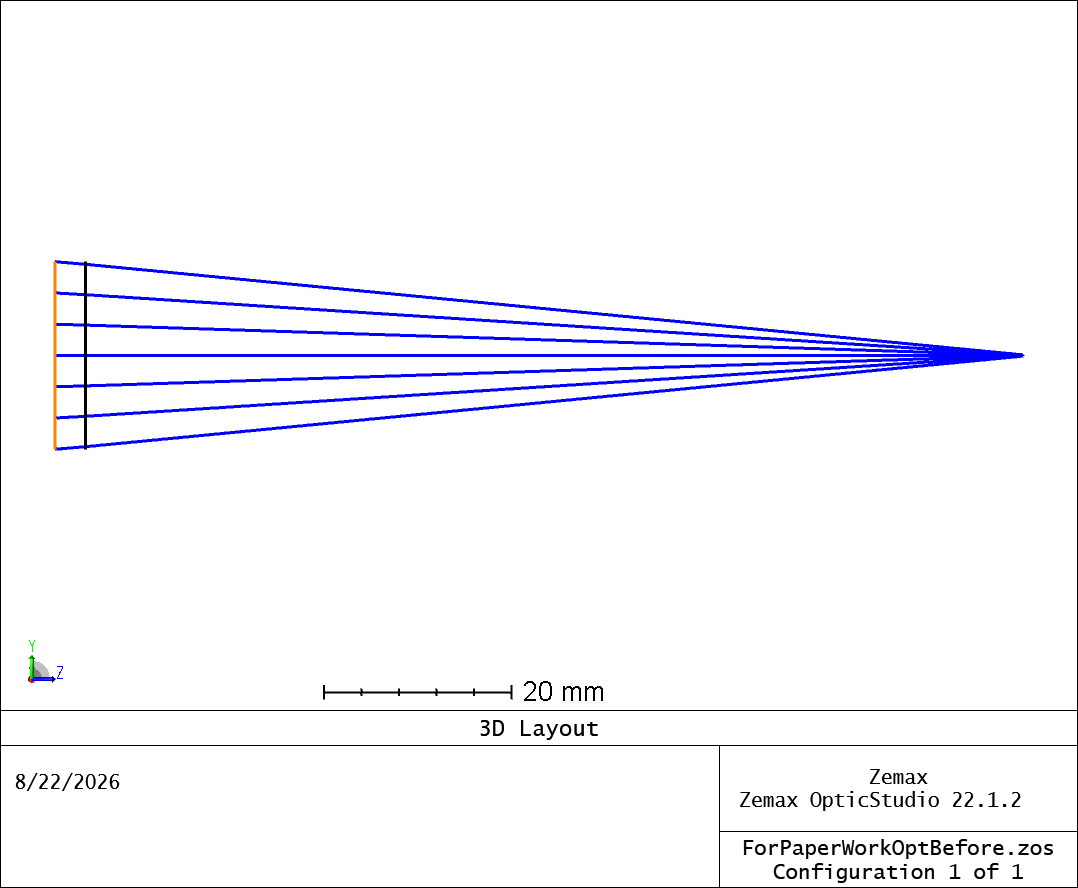}
    \caption{The modified initial configuration composed of two thin lenses (i.e., ``Paraxial'' surfaces in the Zemax OpticStudio). This is equivalent to the configuration of Fig. \ref{fig:e1} in the context of the paraxial approximation (i.e., ABCD matrix).
    The first and second thin lenses have the focal lengths of 118.497 mm and 797.247 mm, respectively. The distance between the first and second thin lenses is 3.293 mm.}
    \label{fig:e3}
\end{figure}
\begin{enumerate}
    \item As preparation, we modify the initial configuration (Fig. \ref{fig:e1}) into the configuration (Fig. \ref{fig:e3}) using only thin lenses equivalent to the original one in the ABCD matrix (paraxial approximation).
    \item We apply the + $\rightarrow$ H transformation to each thin lens in Fig. \ref{fig:e3}, assuming tentative values for the parameters $a$ and $c$ in the + $\rightarrow$ H transformation as initial values of the variables used for the next-step optimization. We implement this step (and preparation of the next step: setting merit function and variables) using a Zemax Programming Language (ZPL) macro.
    \item We execute the optimization (``Hammer Current'' global optimizer in the GUI tools of the Zemax OpticStudio, ``Orthogonal Descent'' algorithm) regarding the parameters $a$ and $c$ in the + $\rightarrow$ H transformation as optimization variables. Here, we use the + $\leftrightarrow$ H transformation as a constraint condition in the merit function. Fig. \ref{fig:e5} shows a screenshot of the optimization tool GUI after the optimization. Fig. \ref{fig:e6}, Table \ref{t2}, and Table \ref{tz2} indicate the resultant configuration of the optimization.
    
\end{enumerate}
The used Zemax files and ZPL macro are available in Ref. \cite{itoh_2026_htransformation_example}.
\begin{figure}[h]
    \centering
    \includegraphics[width=1.0\linewidth]{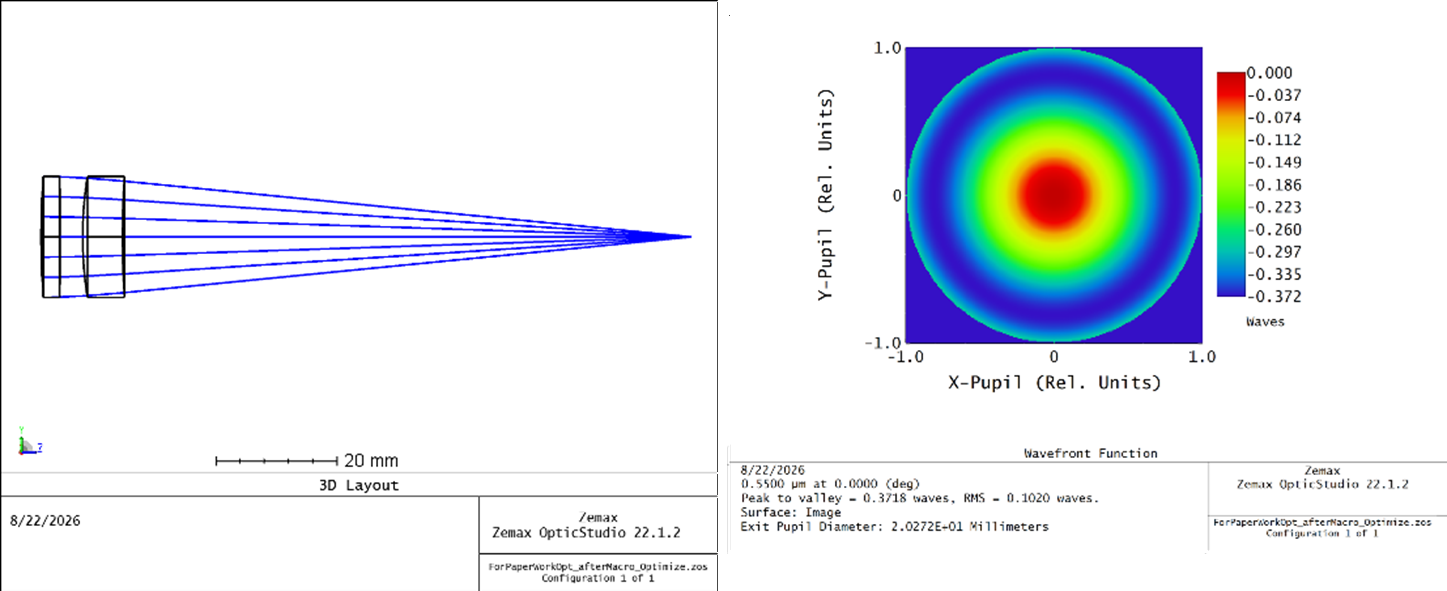}
    \caption{The resultant configuration after the + $\leftrightarrow$ H transformation for the two thin lenses in Fig. \ref{fig:e3}. The left panel shows the optical layout. The right panel shows the wavefront error of this configuration in the case of collimated light injection with the same direction as the optical axis. The assumed wavelenghth of light is $0.55\mathrm{\mu m}$.}
    \label{fig:e6}
\end{figure}

\begin{table}[h]
\caption{The surface parameters of the configuration modified with the + $\leftrightarrow$ H transformation (Fig. \ref{fig:e6}). The second column shows the physical thickness between the surface and the next surface. The third column indicates the refractive index of the medium filling the space between the surface and the next surface.}
  \label{tab:shape-functions}
  \centering
\begin{tabular}{cccc}
\hline
Surface \# &Curvature Radius&Thickness& refractive index \\
\hline
1&123.141 mm&3.122 mm&1.5185\\
2&-792.663 mm&3.755 mm&1.000\\
3&64.230 mm&6.575 mm&1.5185\\
4&161.593 mm&93.845 mm&1.000\\
\hline
\end{tabular}
\label{t2}
\end{table}

\begin{table}[h]
\caption{The Zernike (standard) coefficients up to the 37th order for the resultant configuration (Fig. \ref{fig:e6}). The symbol $\rho$ indicates the pupil radial coordinate normalized by the physical radius of the pupil. The symbol $\lambda$ means the wavelenghth of light; $\lambda=0.55\mathrm{\mu m}$ here. We omitted the orders with the exactly zero values of the coefficients.}
  \label{tab:shape-functions}
  \centering
\begin{tabular}{ccc}
\hline
Order&Base Function&Value of Coefficient\\
\hline
4&$\sqrt{3}(2\rho^2-1)$&-0.0807$\lambda$\\
11&$\sqrt{5}(6\rho^4-6\rho^2+1)$&0.0622$\lambda$\\
22&$\sqrt{7}(20\rho^6-30\rho^4+12\rho^2-1)$&0.0001$\lambda$\\
37&$\sqrt{9}(70\rho^8-140\rho^6+90\rho^4-20\rho^2+1)$&0.0000$\lambda$\\
\hline
\end{tabular}
\label{tz2}
\end{table}
\section{Conclusion\label{secC}}
In this study, we have focused on two fundamental kinds of three-matrix decomposition of the $\mathrm{SL}(2,\mathbb{R})$ matrices (ABCD matrices).
The decomposition of the first kind (+ decomposition) has the form of $T(a)R(\beta)T(c)$, where the symbols $T(p)$ and $R(q)$ denote the transition and refraction matrices defined in Eq. (\ref{eq3}) and Eq. (\ref{eq4}).
We can use the + decomposition for an ABCD matrix defined in Eq. (\ref{eq6}) when $B\neq 0$ or when the ABCD matrix is a transition matrix.
The decomposition of the second kind (H decomposition) has the form of $ R(\alpha)T(b)R(\gamma)$.
The H decomposition is definable when $C\neq 0$ or when the ABCD matrix is a refraction matrix.
When $B=C=0$, the ABCD matrix is decomposable with no + or H decomposition, but is decomposable into two matrices decomposable with + and H decomposition.
Considering the case when $BC\neq 0$ (i.e., the case when the ABCD matrix is decomposable with + and H decomposition), we have formulated a transformation between the parameters in the + and H decompositions (+ $\leftrightarrow$ H transformation) with the definition: $T(a)R(\beta)T(c)= R(\alpha)T(b)R(\gamma)$.
We have found that the + $\leftrightarrow$ H transformation has two fundamental invariants (we define the term “invariant” to refer to a function of a given variable whose functional form (not only the value) remains unchanged even when subjected to a specified variable transformation), which determine the values of the parameter products $\beta b$, $a \alpha$, and $c\gamma$.
We can use the + $\leftrightarrow$ H transformation to increase or decrease the number of refractive surfaces in the optical configuration while keeping the paraxial specifications. This approach is useful in multi-component optical design to optimize non-paraxial performances, such as optical aberrations.

\begin{backmatter}
\bmsection{Disclosures}
The authors declare no conflicts of interest.
\bmsection{Acknowledgment}
This work was supported by JSPS KAKENHI Grant Number JP25K01040.
\bmsection{Data availability} Data underlying the results presented in this paper are available in Ref. \cite{itoh_2026_htransformation_example}.
\end{backmatter}

\bibliography{sample}






\end{document}